\documentclass[aps,prd,preprint,nofootinbib,11pt]{revtex4}
\usepackage{amsfonts}
\usepackage{mathrsfs}
\usepackage{graphicx}
\usepackage{amsmath}
\usepackage{amssymb}
\usepackage{subfigure}
\usepackage{epsfig}
\usepackage{graphicx}
\usepackage{ulem}
\usepackage{color}
\newcommand{\bqa}{\begin{eqnarray}}
\newcommand{\eqa}{\end{eqnarray}}
\newcommand{\beq}{\begin{equation}}
\newcommand{\eeq}{\end{equation}}
\allowdisplaybreaks[1]
\graphicspath{{fig/}{dia/}} \DeclareGraphicsExtensions{.eps}

\begin{document}
\title{\Large Spectroscopy of hidden-heavy tetraquark states with $J^{PC}=0^{--}$ in a color-octet configuration\\[7mm]}

\author{Bing-Dong Wan$^{1,2}$\footnote{wanbd@lnnu.edu.cn}, Jun-Hao Zhang$^{1,2}$, Yan Zhang$^{1,2}$, and Ming-Yang Yuan$^{1,2}$ \vspace{+3pt}}

\affiliation{$^1$Department of Physics, Liaoning Normal University, Dalian 116029, China\\
$^2$Center for Theoretical and Experimental High Energy Physics, Liaoning Normal University, Dalian 116029, China
}

\author{~\\~\\}

\begin{abstract}
\vspace{0.3cm}
Within the QCD sum-rule framework, we investigate hidden-heavy tetraquark channels with the exotic quantum number $J^{PC}=0^{--}$ using four representative local color-octet--octet interpolating currents. The currents include both vector--axialvector and scalar--pseudoscalar Dirac structures. The operator product expansion is carried out up to dimension-eight condensates. The four diagonal sum rules yield mutually consistent mass estimates in the range $10.8$--$11.1~\mathrm{GeV}$ for the hidden-bottom sector and around $4.3$--$4.6~\mathrm{GeV}$ for the corresponding hidden-charm sector, with the bottom sector exhibiting the clearest Borel stability. Since local tetraquark currents with the same quantum numbers are related by Fierz rearrangements, the current-dependent results do not by themselves imply four distinct states or uniquely defined internal color structures. We also discuss quantum-number-allowed decay channels and emphasize the absence of the lowest pseudoscalar--pseudoscalar heavy-meson modes for a neutral $0^{--}$ state. The results provide theoretical guidance for future experimental searches at Belle II, LHCb, and BESIII.
 \end{abstract}
 \maketitle
\newpage

\section{Introduction}

Since the observation of the $X(3872)$ by the Belle Collaboration in 2003~\cite{Choi:2003ue}, heavy-hadron spectroscopy has developed into a broad field in which many candidates for nonconventional matter have been reported~\cite{Chen:2016qju,Liu:2019zoy,Esposito:2016noz,Guo:2017jvc,Brambilla:2019esw,Chen:2022asf,Wang:2025sic,Zhang:2025qmg}. Although the experimental spectrum has become increasingly rich, the internal color organization of these states is still not fully settled. In particular, it remains unclear to what extent a given structure is dominated by color-singlet hadronic degrees of freedom, compact multiquark correlations, or more complicated mixtures of different color components.

This issue becomes especially sharp in channels with manifestly exotic quantum numbers. Among them, $J^{PC}=0^{--}$ is particularly attractive because such a quantum number cannot be generated by an ordinary quark--antiquark meson. Any convincing signal in this channel would therefore point directly to genuinely exotic QCD dynamics. For this reason, the $0^{--}$ sector has long been regarded as a useful testing ground for unconventional hadronic configurations, including glueballs, hybrids, and multiquark systems~\cite{Qiao:2014vva,Qiao:2015iea,Wan:2023epq,Tang:2025ept,Huang:2016rro}.

In hidden-heavy systems, one possible realization of a $0^{--}$ state is through the coupling of two mesonlike subclusters. Earlier analyses have mainly emphasized molecular configurations built from color-singlet pairs. However, QCD also allows a different clustering pattern, namely an octet--octet combination of the schematic form $[q\bar{q}]_{8_c}\otimes[q\bar{q}]_{8_c}$. The octet--octet notation used here refers to a color-coupling basis of an overall color-singlet local four-quark operator. It does not imply the existence of two independently bound color-octet mesons. At short distances, the one-gluon-exchange interaction within an isolated $q\bar q$ color octet is repulsive. However, such an isolated colored subsystem is not a physical asymptotic state, and the complete color-singlet four-quark system also contains cross-pair color interactions between quarks belonging to different pairings. For an overall color-singlet four-quark system,
\begin{equation}
\sum_{i<j}\boldsymbol{T}_i\cdot\boldsymbol{T}_j
=
\frac{1}{2}
\left(
C_{\rm total}-\sum_i C_i
\right)
=
-\frac{8}{3}.
\end{equation}
In an $8\otimes 8\to 1$ coupling, the two selected $q\bar q$ pairs contribute $1/6+1/6=1/3$, so that the sum of the remaining cross-pair color factors is $-3$. This color-algebra result shows that the repulsion inside the two selected octet pairings does not by itself exclude an attractive component in the complete color-singlet four-quark system~\cite{Tang:2019nwv,Tang:2016pcf,Dubynskiy:2008mq,Tang:2024zvf,Tang:2024kmh}. It only clarifies that the present QCD sum rule probes the full color-singlet correlator and does not assume two separately bound octet subclusters. Throughout this work, the color-octet notation refers to the color-coupling basis of the interpolating currents and not to a uniquely determined physical substructure.

Motivated by this possibility, we investigate hidden-heavy tetraquark channels with $J^{PC}=0^{--}$ using color-octet--octet interpolating currents. The central question of the present work is whether local color-octet--octet interpolating currents generate stable QCD sum rules for low-lying hidden-heavy $0^{--}$ spectral strength, rather than whether two isolated color-octet subclusters form independently bound objects. This issue is particularly important in the bottom sector, where a cleaner OPE hierarchy and a more favorable pole contribution make it possible to test the stability of the sum rules constructed from the color-octet--octet currents more stringently than in the corresponding charm channel.

In the present work, we construct interpolating currents for the configurations $[Q\bar{q}]_{8_c}\otimes[q\bar{Q}]_{8_c}$ and $[Q\bar{Q}]_{8_c}\otimes[q\bar{q}]_{8_c}$ with $Q=c,b$ and analyze the corresponding two-point correlation functions within the QCD sum rule framework~\cite{Shifman:1978bx,Reinders:1984sr}. The operator product expansion is carried out up to dimension-eight condensates, and the resulting sum rules are used to extract the masses of the lowest-lying states. In this way, the present study extends the spectroscopy of hidden-heavy $0^{--}$ systems from the conventional molecular picture~\cite{Wan:2024fam} to a color-octet--octet framework and allows a systematic comparison among different octet pairing patterns. QCD sum rules have been widely employed in hadron spectroscopy and remain a valuable nonperturbative tool for exploring both conventional and exotic configurations~\cite{Shifman:1978bx,Reinders:1984sr,Albuquerque:2013ija,P.Col,Narison:1989aq,Govaerts:1984hc,Wang:2013vex,Wan:2019ake,Tang:2021zti,Wan:2020oxt,Wan:2020fsk,Wang:2020cme,Yang:2020wkh,Wang:2021qmn,Wan:2021vny,Yin:2021cbb,Wan:2022xkx,Zhang:2022obn,Wan:2022uie,Agaev:2022pis,Zhao:2023imq,Li:2024ctd,Tang:2024zvf,Wan:2024dmi,Wan:2024fam,Wan:2024pet,Wan:2024ykm,Zhang:2024ick,Zhang:2024jvv,Zhang:2024ulk,Zhang:2024asb,Tang:2024kmh,Wan:2025xhf,Wan:2025bdr,Wan:2025zau,Agaev:2025llz,Barsbay:2025vjq,Wan:2025fyj,Zhang:2025vqg,Wan:2025sae,Wan:2025ikc,Mu:2026bue,Ben:2025wqn}.

The paper is organized as follows. In Sec.\ref{Formalism}, we present the QCD sum rule formalism and construct the interpolating currents with the color-octet--octet coupling basis. The numerical analysis and the extraction of the masses are performed in Sec.\ref{Numerical}. In Sec.\ref{Decay}, we analyze the possible decay channels. Finally, a brief summary is given in Sec.\ref{Summary}.

\section{Formalism}\label{Formalism}

The analysis begins with the construction of local operators carrying $J^{PC}=0^{--}$ and written in a color-octet--octet coupling basis. For the present purpose, we consider four representative local interpolating currents in the selected color-octet--octet operator basis. These currents can be arranged into two categories: hidden-flavor octet--octet pairing, $[\bar{Q}Q]_{\mathbf{8}}\otimes[\bar{q}q]_{\mathbf{8}}$, and open-flavor octet--octet pairing, $[\bar{Q}q]_{\mathbf{8}}\otimes[\bar{q}Q]_{\mathbf{8}}$. Their explicit forms are given by:
\begin{eqnarray}
J_A(x) &=& i[\bar{Q}_i \gamma_\mu (t^a)_{ij} Q_j][\bar{q}_m \gamma^\mu\gamma_5 (t^a)_{mn} q_n] \;, \\
J_B(x) &=& i[\bar{Q}_i \gamma_\mu\gamma_5 (t^a)_{ij} Q_j][\bar{q}_m \gamma^\mu (t^a)_{mn} q_n] \;, \\
J_C(x) &=& \frac{i}{\sqrt{2}}\Big( [\bar{Q}_i \gamma_\mu (t^a)_{ij} q_j][\bar{q}_m \gamma^\mu\gamma_5 (t^a)_{mn} Q_n] \nonumber \\
&& + [\bar{Q}_i \gamma_\mu\gamma_5 (t^a)_{ij} q_j][\bar{q}_m \gamma^\mu (t^a)_{mn} Q_n] \Big) \;, \\
J_D(x) &=& \frac{i}{\sqrt{2}}\Big( [\bar{Q}_i \gamma_5 (t^a)_{ij} q_j][\bar{q}_m (t^a)_{mn} Q_n] \nonumber \\
&& - [\bar{Q}_i (t^a)_{ij} q_j][\bar{q}_m \gamma_5 (t^a)_{mn} Q_n] \Big) \;.
\end{eqnarray}
Here, $Q$ denotes the heavy quark $c$ or $b$, $q$ stands for the light quark $u$ or $d$, and $t^a=\lambda^a/2$ are the generators of $SU(3)_c$. The color indices are labeled by $i$, $j$, $m$, and $n$. In what follows, the four operators introduced above will be denoted by $J_A$, $J_B$, $J_C$, and $J_D$.

The four currents contain two different classes of Dirac structures. The currents $J_A$, $J_B$, and $J_C$ are of vector--axialvector type, whereas $J_D$ is a scalar--pseudoscalar current. In particular, $J_D$ is constructed from the antisymmetric combination $P\otimes S-S\otimes P$ in the open-flavor pairing. Under parity and charge conjugation, the current $J_D$ transforms as
\begin{equation}
\mathcal{P}J_D(t,\boldsymbol{x})\mathcal{P}^{-1}
=-J_D(t,-\boldsymbol{x}),
\end{equation}
and
\begin{equation}
\mathcal{C}J_D(x)\mathcal{C}^{-1}
=-J_D(x),
\end{equation}
respectively. Therefore, $J_D$ carries $J^{PC}=0^{--}$. A hidden-flavor local product such as $(\bar Q t^a Q)(\bar q i\gamma_5 t^a q)$ is even under charge conjugation and therefore carries $0^{-+}$ rather than $0^{--}$. The open-flavor antisymmetric scalar--pseudoscalar combination used in $J_D$ is required to obtain $C=-$ without introducing derivatives.

\begin{table}[t]
\caption{Flavor arrangement and Dirac structures of the interpolating currents.}
\centering
\begin{tabular}{c c c c}
\hline\hline
Current & Flavor arrangement & Dirac structure & $J^{PC}$ \\
\hline
$J_A$ & $[\bar QQ]_{8_c}[\bar qq]_{8_c}$ & $V\otimes A$ & $0^{--}$ \\
$J_B$ & $[\bar QQ]_{8_c}[\bar qq]_{8_c}$ & $A\otimes V$ & $0^{--}$ \\
$J_C$ & $[\bar Qq]_{8_c}[\bar qQ]_{8_c}$ &
$V\otimes A+A\otimes V$ & $0^{--}$ \\
$J_D$ & $[\bar Qq]_{8_c}[\bar qQ]_{8_c}$ &
$P\otimes S-S\otimes P$ & $0^{--}$ \\
\hline\hline
\end{tabular}
\end{table}

The octet--octet notation specifies the color coupling of the local operator and should not be interpreted as a unique decomposition of the physical state. Local tetraquark operators carrying the same quantum numbers can be related by color and Dirac Fierz rearrangements. For example, the color identity
\begin{equation}
(t^a)_{ij}(t^a)_{mn}
=
\frac{1}{2}\delta_{in}\delta_{mj}
-\frac{1}{6}\delta_{ij}\delta_{mn}
\end{equation}
shows explicitly that an octet--octet operator can be rewritten as a linear combination of operators with different color pairings. A subsequent Dirac Fierz transformation relates these operators to meson--meson-type and diquark--antidiquark-type local currents. Accordingly, the current structure specifies the operator used to probe the spectrum, rather than uniquely determining the spatial or color wave-function decomposition of the physical state.

\begin{figure}[htb]
\centering
\includegraphics[width=10cm]{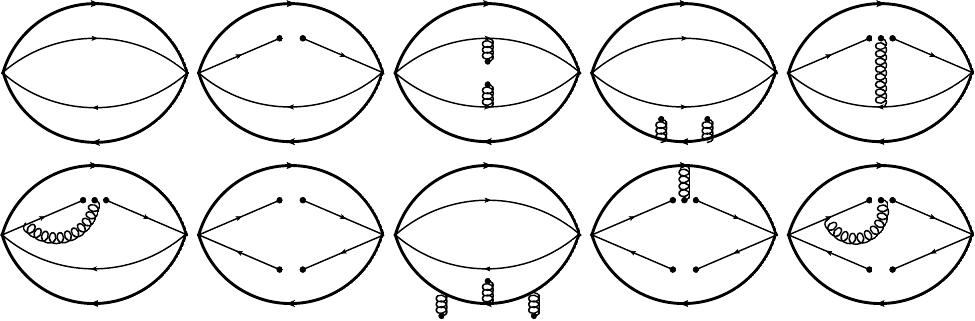}
\caption{Feynman diagrams contributing to the OPE calculation. The thick solid line represents the heavy quark, the thin solid line corresponds to the light quark, and the coiled line represents the gluon. Contributions from heavy quark condensates are suppressed in the large mass limit.} \label{feyndiag}
\end{figure}

Once the interpolating currents are specified, the corresponding hadronic information is encoded in the two-point correlation function
\begin{eqnarray}
\Pi(q^2) &=& i \int d^4 x\, e^{i q \cdot x} \langle 0 | T \{ J(x)\, J^\dagger(0) \} |0 \rangle \;,
\end{eqnarray}
which connects the quark--gluon description to the hadronic spectrum. In the deep Euclidean region, the correlation function is expanded in local operators. In the present formulation, the OPE contributions relevant to the sum rules are organized directly into the spectral representation, so that the working expression takes the form
\begin{eqnarray}
\Pi^{\mathrm{OPE}}(q^2)
&=& \int_{(2m_Q)^2}^{\infty} ds\, \frac{\rho^{\mathrm{OPE}}(s)}{s-q^2}  \;.
\end{eqnarray}
In the present calculation, the spectral density contains the perturbative term together with condensate contributions up to dimension eight. For brevity, the condensate terms entering the working sum rules are collectively absorbed into the definition of $\rho^{\mathrm{OPE}}(s)$ in the present presentation,
\begin{eqnarray}
\rho^{\mathrm{OPE}}(s) &=& \rho^{\mathrm{pert}}(s) + \rho^{\langle \bar{q}q \rangle}(s) + \rho^{\langle G^2 \rangle}(s) + \dots + \rho^{\langle \bar{q}q \rangle \langle \bar{q}Gq \rangle}(s) \;.
\end{eqnarray}
The typical Feynman diagrams contributing to these terms are shown in Fig.~\ref{feyndiag}. To keep the presentation concise, we do not list the lengthy analytical expressions for the spectral densities in this paper; they can be provided by the authors upon reasonable request.

On the hadronic side, the lowest pole is separated from the higher resonances and continuum states in the standard way,
\begin{eqnarray}
\Pi^{\mathrm{phen}}(q^2)
&=& \frac{\lambda_H^2}{M_H^2 - q^2}
+ \int_{s_0}^{\infty} ds\, \frac{\rho^{\mathrm{cont}}(s)}{s - q^2} \;,
\end{eqnarray}
where $M_H$ denotes the tetraquark mass and $\lambda_H$ is the corresponding current--hadron coupling.

After invoking quark--hadron duality and applying the Borel transformation $\mathcal{B}_{M_B^2}$ to suppress the continuum contribution, one arrives at the working sum rules. The hadron mass is extracted from the ratio of moments,
\begin{eqnarray}
M_H(s_0, M_B^2) &=& \sqrt{\frac{\mathcal{L}_1(s_0, M_B^2)}{\mathcal{L}_0(s_0, M_B^2)}} \;,
\end{eqnarray}
where
\begin{eqnarray}
\mathcal{L}_0(s_0, M_B^2) &=& \int_{(2m_Q)^2}^{s_0} ds\, \rho^{\mathrm{OPE}}(s) e^{-s/M_B^2}  \;, \\
\mathcal{L}_1(s_0, M_B^2) &=& \frac{\partial}{\partial \left( -1/M_B^2 \right)} \mathcal{L}_0(s_0, M_B^2) \;.
\end{eqnarray}
The Borel parameter $M_B^2$ improves the convergence of the truncated OPE and at the same time reduces the contamination from excited states.

\section{Numerical analysis}\label{Numerical}

The numerical evaluation is performed with the standard set of QCD input parameters~\cite{Albuquerque:2013ija,P.Col,Tang:2019nwv,Narison:1989aq,Govaerts:1984hc}: $m_c(m_c)=\overline{m}_c=(1.275\pm0.025)\,\text{GeV}$, $m_b(m_b)=\overline{m}_b=(4.18\pm0.03)\,\text{GeV}$, $\langle \bar{q} q \rangle = -(0.23\pm0.03)^3\,\text{GeV}^3$, $\langle \bar{q} g_s \sigma \cdot G q \rangle = m_0^2 \langle\bar{q} q \rangle$, $\langle g_s^2 G^2 \rangle = (0.88\pm0.25)\,\text{GeV}^4$, $\langle g_s^3 G^3 \rangle = (0.045\pm0.013)\,\text{GeV}^6$, and $m_0^2=(0.8\pm0.2)\,\text{GeV}^2$. Throughout this work, the light-quark masses are neglected, $m_u=m_d=0$.

For each current, the admissible sum-rule window is determined by balancing OPE convergence against pole dominance. The first requirement is that the highest-dimension contribution retained in the truncated expansion remain numerically under control. To monitor this condition, we define
\begin{eqnarray}
R^{\mathrm{OPE}}=\left|\frac{L_0^{\mathrm{dim}=8}(s_0,M_B^2)}{L_0(s_0,M_B^2)}\right| \; .
\end{eqnarray}
In the accepted Borel windows, the dimension-eight contribution remains numerically subleading compared with the total OPE contribution. In practice, we require this contribution to remain at a modest level~\cite{P.Col}. The second requirement is that the lowest hadronic pole still provide a substantial fraction of the full sum rule,
\begin{eqnarray}
R^{\mathrm{PC}}=\frac{L_0(s_0,M_B^2)}{L_0(\infty,M_B^2)} \; . \label{RatioPC}
\end{eqnarray}
For the pole contribution to possess a physical interpretation as a fraction of the total sum rule, we additionally require
\begin{equation}
0<\mathcal{L}_0(s_0,M_B^2)
<
\mathcal{L}_0(\infty,M_B^2),
\end{equation}
which ensures
\begin{equation}
0<R^{\rm PC}(s_0,M_B^2)<1.
\end{equation}
Together with the pole-dominance condition $R^{\rm PC}\geq 50\%$, the accepted range is therefore
\begin{equation}
0.5\leq R^{\rm PC}<1.
\end{equation}
In practice, $s_0$ is determined together with the Borel window. For each current, we scan $\sqrt{s_0}$ and require that the OPE-convergence and pole-dominance criteria be simultaneously satisfied. The lower boundary of the Borel window is determined by the more restrictive of the OPE-convergence condition and the requirement $R^{\rm PC}<1$, while the upper boundary is fixed by the conventional pole-dominance criterion $R^{\rm PC}\geq 50\%$. We then choose the value of $s_0$ for which the extracted mass shows the weakest dependence on $M_B^2$ within the accepted window. The uncertainty from the continuum threshold is estimated by varying $\sqrt{s_0}$ by about $0.1~\mathrm{GeV}$ around its central value~\cite{Wan:2020oxt,Wan:2020fsk,Wan:2019ake}.

\begin{figure}
\includegraphics[width=6.8cm]{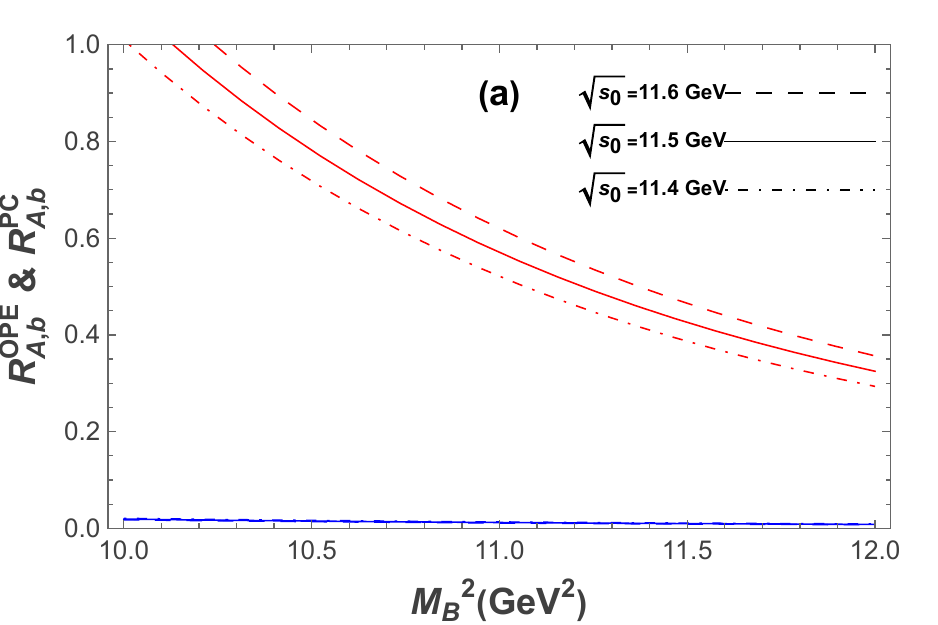}
\includegraphics[width=6.8cm]{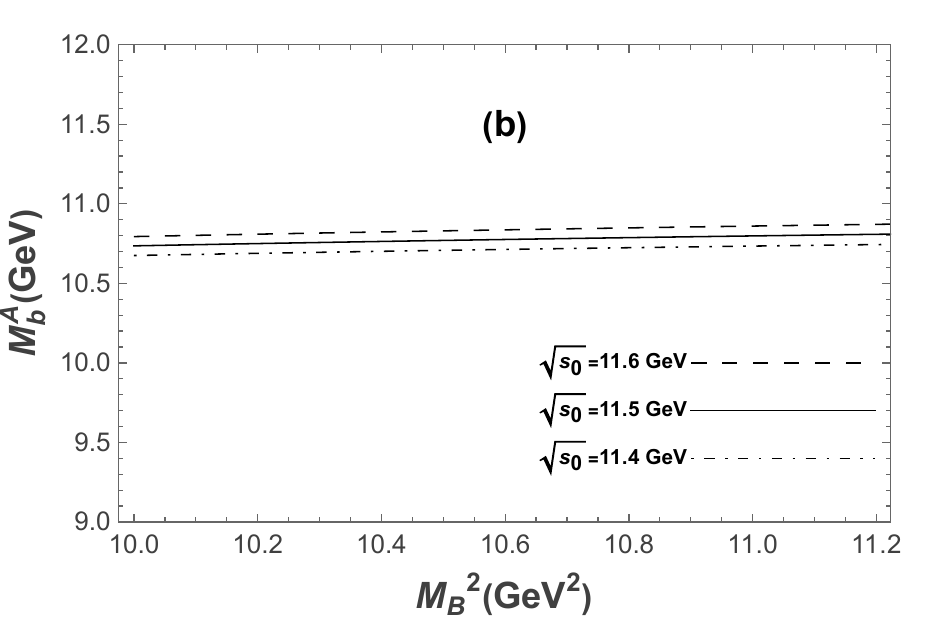}
\caption{(a) The ratios of ${R_{A}^{OPE}}$ and ${R_{A}^{PC}}$ as functions of the Borel parameter $M_B^2$ for different values of $\sqrt{s_0}$, where blue lines represent ${R_{A}^{OPE}}$ and red lines denote ${R_{A}^{PC}}$. (b) The mass $M_{b}^{A}$ as a function of the Borel parameter $M_B^2$ for different values of $\sqrt{s_0}$. The curves are displayed over ranges wider than the adopted fiducial Borel windows. Only the regions satisfying the OPE-convergence condition and $0.5\leq R^{\rm PC}<1$ are used in the numerical extraction. At smaller $M_B^2$, the finite-order truncated OPE does not preserve the positivity relation required for a physical pole-fraction interpretation; these points are outside the accepted windows and do not enter the mass analysis.} \label{figAb0--}
\end{figure}

Among the channels studied here, the hidden-bottom sector exhibits the cleanest sum-rule behavior and is therefore used as the primary benchmark for the color-octet--octet scenario. For the representative current $J_A$, the OPE convergence, pole contribution, and mass curve are displayed in Fig.~\ref{figAb0--}. The pole-contribution curves in Figs.~2--9 are displayed over Borel ranges wider than the final fiducial windows in order to show their overall behavior. The exact hadronic spectral density is non-negative. However, the quantity entering the ratio in Eq.~(\ref{RatioPC}) is the finite-order truncated OPE spectral density, including condensate contributions, and it is not guaranteed to preserve pointwise positivity outside the fiducial sum-rule region. At sufficiently small $M_B^2$, the truncated OPE may therefore violate the positivity relation required for interpreting $R^{\rm PC}$ as a physical fraction. Such regions are excluded by the condition $0<R^{\rm PC}<1$ and are not used in the numerical extraction. As can be seen from the figure, the contribution of the highest-dimension term remains comparatively small in the adopted window, while the pole contribution stays sufficiently sizable. At $\sqrt{s_0}=11.5$ GeV, the lower edge of the Borel window is determined by the more restrictive of the OPE criterion and the condition $R^{\rm PC}<1$, whereas the upper edge is constrained by the pole-dominance requirement. In this way we obtain
\begin{eqnarray}
10.3 \le M_B^2 \le 11.2~\text{GeV}^2 \; ,
\end{eqnarray}
from which the hidden-bottom mass is extracted as
\begin{eqnarray}
M_{b}^{A}=(10.78\pm0.09)\,\text{GeV} \; .
\end{eqnarray}
This result indicates that the correlator constructed from $J_A$ supports a stable low-lying pole solution in this mass region.

The corresponding hidden-charm partner is analyzed in the same manner. Using the same current $J_A$, we obtain
\begin{eqnarray}
M_{c}^{A}=(4.38\pm0.13)\,\text{GeV} \; .
\end{eqnarray}
with the working interval
\begin{eqnarray}
2.5 \le M_B^2 \le 3.1~\text{GeV}^2 \; ,
\end{eqnarray}
for $\sqrt{s_0}=5.1$ GeV. Compared with the bottom sector, the charm channel is more sensitive to the sum-rule parameters, but it still provides a useful companion result for comparison with the hidden-bottom analysis.

The other three currents lead to a similar overall pattern. In the hidden-bottom sector, the extracted masses are $(10.82\pm0.08)$ GeV for $J_B$, $(11.08\pm0.08)$ GeV for $J_C$, and $(10.90\pm0.08)$ GeV for $J_D$, with the corresponding working windows $10.0\le M_B^2\le10.9~\text{GeV}^2$, $10.6\le M_B^2\le11.6~\text{GeV}^2$, and $10.3\le M_B^2\le11.2~\text{GeV}^2$, respectively. In the hidden-charm sector, the extracted masses are $(4.30\pm0.15)$ GeV for $J_B$, $(4.57\pm0.11)$ GeV for $J_C$, and $(4.46\pm0.11)$ GeV for $J_D$, with the corresponding Borel windows $2.1\le M_B^2\le2.7~\text{GeV}^2$, $2.6\le M_B^2\le3.3~\text{GeV}^2$, and $2.3\le M_B^2\le2.9~\text{GeV}^2$, respectively. For these currents, the corresponding OPE and pole-contribution curves exhibit the same qualitative balance used to define the representative $J_A$ window.

The moderate spread among the current-dependent mass estimates should not be directly interpreted as evidence for several distinct internal clustering structures. Since local currents with the same quantum numbers are related by Fierz rearrangements, different currents may have nonzero overlap with the same physical state. The observed spread therefore mainly reflects the current dependence and the systematic uncertainty of the truncated diagonal sum-rule analysis. Since only diagonal correlation functions are considered in the present work, we cannot determine whether the four currents predominantly couple to distinct states or to the same low-lying state with different residues. A correlation-matrix analysis including off-diagonal correlators would be required to resolve this issue.

It is also interesting to note that the hidden-charm masses extracted from the open-flavor-type currents $J_C$ and $J_D$ lie close to the mass region predicted for the $D_1\bar D^*$ and $D_{s1}\bar D_s^*$ molecular candidates in previous studies~\cite{Peng:2022nrj}. The proximity of some current-dependent mass estimates to open-heavy thresholds may indicate potentially important coupled hadronic channels. However, this numerical proximity alone does not determine the internal structure of the state. A quantitative analysis of current mixing and off-diagonal correlators is beyond the scope of the present work.

It is worth emphasizing that both the hidden-bottom and hidden-charm solutions populate relatively compact spectral regions. The hidden-bottom masses lie in the range from about $10.8$ to $11.1~\mathrm{GeV}$, while the corresponding hidden-charm partners are found around $4.3$--$4.6~\mathrm{GeV}$. The main difference between the two sectors is therefore not the size of the mass spread, but the stability of the extracted masses with respect to the Borel parameter $M_B^2$. In the hidden-bottom sector, the mass curves exhibit flatter Borel platforms, indicating a weaker dependence on $M_B^2$. By contrast, the hidden-charm sector is more sensitive to the choice of the Borel window and the continuum threshold. The four diagonal sum rules consistently indicate a low-lying spectral region. The corresponding OPE and pole-contribution curves for $J_B$-$J_D$ show analogous stability behavior and are displayed in Appendix~\ref{pictures}.

For comparison, the outputs for all four currents in both the hidden-bottom and hidden-charm sectors are summarized in Table~\ref{mass}. In the present work, all four currents are treated as completed numerical results, with $J_A$ shown as a representative example in the main text and the corresponding OPE, pole-contribution, and mass-stability curves for the other channels collected in Appendix~\ref{pictures}. The quoted uncertainties mainly arise from the variations of the heavy-quark masses, condensates, and continuum thresholds, among which the continuum threshold and heavy-quark masses provide the dominant sources of variation.

\begin{table}
\begin{center}
\renewcommand\tabcolsep{10pt}
\caption{The continuum thresholds, Borel windows, pole-contribution ranges, and predicted masses of hidden-charm and hidden-bottom tetraquark channels. The quoted $R^{\rm PC}$ ranges correspond to the central values of $\sqrt{s_0}$ and are evaluated over the adopted Borel windows.}\label{mass}
\begin{tabular}{cccccc}\hline\hline
                 &Current   & $\sqrt{s_0}\;(\text{GeV})$     &$M_B^2\;(\text{GeV}^2)$ &$R^{\rm PC}$ range &$M^X\;(\text{GeV})$       \\ \hline
$b$-sector      &$A$        & $(11.5\pm0.1)$                           &$10.3-11.2$                  &$0.50-0.89$&$(10.78\pm0.09)$           \\
                        &$B$        & $(11.5\pm0.1)$                           &$10.0-10.9$                  &$0.50-0.90$&$(10.82\pm0.08)$          \\      
                        &$C$        & $(11.9\pm0.1)$                            &$10.6-11.6$                  &$0.50-0.91$&$(11.08\pm0.08)$          \\  
                       &$D$        & $(11.6\pm0.1)$                            &$10.3-11.2$                    &$0.50-0.89$&$(10.90\pm0.08)$          \\\hline
$c$-sector      &$A$        & $(5.1\pm0.1)$                             &$2.5-3.1$                      &$0.50-0.75$&$(4.38\pm0.13)$         \\
                        &$B$        & $(4.9\pm0.1)$                             &$2.1-2.7$                      &$0.50-0.88$&$(4.30\pm0.15)$          \\
                        &$C$        & $(5.3\pm0.1)$                             &$2.6-3.3$                      &$0.50-0.71$&$(4.57\pm0.11)$ \\
                        &$D$        & $(5.1\pm0.1)$                             &$2.3-2.9$                      &$0.50-0.80$&$(4.46\pm0.11)$       \\
\hline
 \hline
\end{tabular}
\end{center}
\end{table}
 
\section{Decay Patterns and Possible Search Channels}
\label{Decay}

The study of decay patterns is essential for the experimental identification of the exotic $J^{PC}=0^{--}$ tetraquark states, especially because this quantum number is forbidden for conventional $q\bar{q}$ mesons. The quantum-number constraints from $J$, $P$, and $C$ conservation therefore provide particularly useful signatures for distinguishing such states from ordinary heavy quarkonia and other charmonium-like or bottomonium-like backgrounds.

A general and important feature of a neutral $0^{--}$ state is the absence of the lowest pseudoscalar--pseudoscalar heavy-meson channels. For a pair of pseudoscalar mesons such as $D\bar D$ or $B\bar B$, the total spin is $S=0$, and hence the total angular momentum is fixed by the relative orbital angular momentum, $J=L$. The parity and charge conjugation are given by
 \[
 P=(-1)^L,\qquad C=(-1)^L .
 \]
Thus, the $J=0$ configuration requires $L=0$, which gives $J^{PC}=0^{++}$ rather than $0^{--}$. Although an odd-$L$ pseudoscalar--pseudoscalar pair has $P=C=-1$, it carries $J=L\neq 0$ and therefore cannot match a $0^{--}$ initial state. Consequently, a neutral $0^{--}$ state cannot decay into the lowest pseudoscalar--pseudoscalar channels such as $D\bar D$ or $B\bar B$. This selection rule provides a clean signature of the exotic $0^{--}$ assignment.

In the hidden-bottom sector, which provides the most stable sum-rule signals in the present work, open-bottom final states are expected to play an important role in experimental searches. The predicted masses, lying around $10.8$--$11.1~\mathrm{GeV}$, are above the lowest open-bottom thresholds, while the pseudoscalar--pseudoscalar channel $B\bar B$ is forbidden by the $0^{--}$ quantum numbers. The relevant low-lying open-bottom modes should therefore involve at least one vector or orbitally excited bottom meson. Depending on the precise mass of the state, possible search channels include $B\bar B^*+\mathrm{c.c.}$ in a $P$ wave, understood as the charge-conjugation eigencombination compatible with the $0^{--}$ assignment. If kinematically allowed, the $S$-wave channel $B^*\bar B_1+\mathrm{c.c.}$, again with the appropriate charge-conjugation combination, may also become important. The latter mode is particularly interesting because it can proceed in an $S$ wave without a centrifugal barrier. However, its actual coupling strength cannot be inferred from the Dirac structure of the two-point interpolating current and requires a dedicated decay calculation. Experimentally, it would therefore be useful to search for such structures in open-bottom invariant-mass distributions, supplemented by hidden-bottom final states with cleaner backgrounds.

For comparison, the hidden-charm partners lie in a mass region where several open-charm thresholds are nearby, so their decay patterns are expected to be sensitive to the precise mass and phase space. As in the hidden-bottom sector, the pseudoscalar--pseudoscalar channel $D\bar D$ is forbidden for a neutral $0^{--}$ state by angular-momentum, parity, and charge-conjugation constraints. The relevant low-lying open-charm modes should therefore involve at least one vector or orbitally excited charmed meson. In the lower part of the predicted mass region, $D\bar D^*+\mathrm{c.c.}$ in a $P$ wave provides a possible search channel, where the charge-conjugation eigencombination should be chosen to match the $0^{--}$ assignment. For candidates close to or above the $D^*\bar D_1$ threshold, the $S$-wave channel $D^*\bar D_1+\mathrm{c.c.}$ may become important, again with the appropriate charge-conjugation combination. This channel is compatible with the required quantum numbers and may be experimentally relevant because it can proceed in an $S$ wave. Its actual coupling strength cannot be inferred from the Dirac structure of the two-point interpolating current and requires a dedicated decay calculation. Hidden-charm final states generated by quark rearrangement, such as charmonium plus light mesons, may also contribute, but their relative strengths require a dedicated decay calculation and cannot be determined from the present mass sum rules. The detailed decay pattern may therefore depend on the mass position relative to nearby thresholds and possible mixing among currents.

In summary, the most characteristic experimental feature of a $0^{--}$ hidden-heavy tetraquark is the absence of the $pseudoscalar$--$pseudoscalar$ channels such as $D\bar D$ and $B\bar B$, together with possible signals in open-flavor modes such as $D\bar D^*$, $D^*\bar D_1$, $B\bar B^*$, and $B^*\bar B_1$, with the charge-conjugation combinations chosen to match the $0^{--}$ assignment. We therefore encourage dedicated searches for such exotic structures in both open-flavor and hidden-flavor final states at Belle~II and LHCb, with complementary searches for the hidden-charm partners at BESIII.

\section{Summary}
\label{Summary}

In this work, we have performed a QCD sum rule study of hidden-heavy tetraquark channels with the exotic quantum number $J^{PC}=0^{--}$ using color-octet--octet interpolating currents. Since the $0^{--}$ assignment is forbidden for conventional $q\bar{q}$ mesons, the possible existence of such states would provide a clean signal of genuinely exotic hadronic matter.

By constructing four representative color-octet--octet interpolating currents (including both vector--axialvector and scalar--pseudoscalar Dirac structures) and evaluating the corresponding two-point correlation functions up to dimension-eight condensates, we obtain a set of hidden-heavy $0^{--}$ mass predictions. The resulting masses are summarized in Table~\ref{mass}. The hidden-bottom sector shows the clearest Borel stability, while the hidden-charm sector provides a useful companion comparison with a stronger sensitivity to the sum-rule parameters. The four diagonal sum rules consistently indicate a low-lying spectral region. Since the currents are related by Fierz rearrangements and the off-diagonal correlators have not been studied, the present analysis does not establish the number of physical states or a unique internal color decomposition. A correlation-matrix analysis including off-diagonal correlators would be required to resolve this issue.

We have also briefly discussed possible decay signatures of these states. For a neutral $0^{--}$ state, the lowest pseudoscalar--pseudoscalar channels, such as $D\bar D$ and $B\bar B$, cannot match the required angular momentum, parity, and charge conjugation and are therefore forbidden. This selection rule provides a useful experimental discriminator for the exotic $0^{--}$ assignment. The relevant low-lying open-flavor modes are expected to involve at least one vector or orbitally excited heavy meson. Possible search channels include $D\bar D^*+\mathrm{c.c.}$ and $B\bar B^*+\mathrm{c.c.}$ in $P$ waves, as well as $D^*\bar D_1+\mathrm{c.c.}$ and $B^*\bar B_1+\mathrm{c.c.}$ in $S$ waves when they are kinematically accessible. The actual coupling strengths of these channels cannot be determined from the present two-point sum rules and require dedicated decay calculations. The present results may therefore provide useful guidance for future searches for hidden-heavy exotic states at Belle~II and LHCb, with complementary probes of the hidden-charm partners at BESIII.

%%%%%%%%%%%%%%%%%%%%%%%%%%%%%%%%%%%%%%%%%%%%%%%%%%%%%%%%%%%%%%%%%%%%%
\vspace{0.5cm} {\bf Acknowledgments}

This work was supported in part by the National Natural Science Foundation of China under Grants 12575106 and 12147214, and Specific Fund of Fundamental Scientific Research Operating Expenses for Undergraduate Universities in Liaoning Province under Grants No. LJ212410165019.
During the preparation of this manuscript, the authors used ChatGPT (OpenAI) only for language polishing. The authors carefully checked and revised the manuscript and take full responsibility for all scientific content, calculations, analyses, interpretations, and conclusions.
%%%%%%%%%%%%%%%%%%%%%%%%%%%%%%%%%%%%%%%%%%%%%%%%%%%%%%%%%%%%%%%%%%%%

\begin{widetext}
\appendix
\section{Supplementary Figures}\label{pictures}
For the hidden-charm and hidden-bottom $0^{--}$ tetraquark states associated with the currents $J_B$-$J_D$, together with the hidden-charm partner of $J_A$, the OPE, pole contribution, and masses as functions of the Borel parameter $M_B^2$ are given in Figs.~\ref{figA0--} to \ref{figDb0--}. 

\begin{figure}[h]
\includegraphics[width=6.8cm]{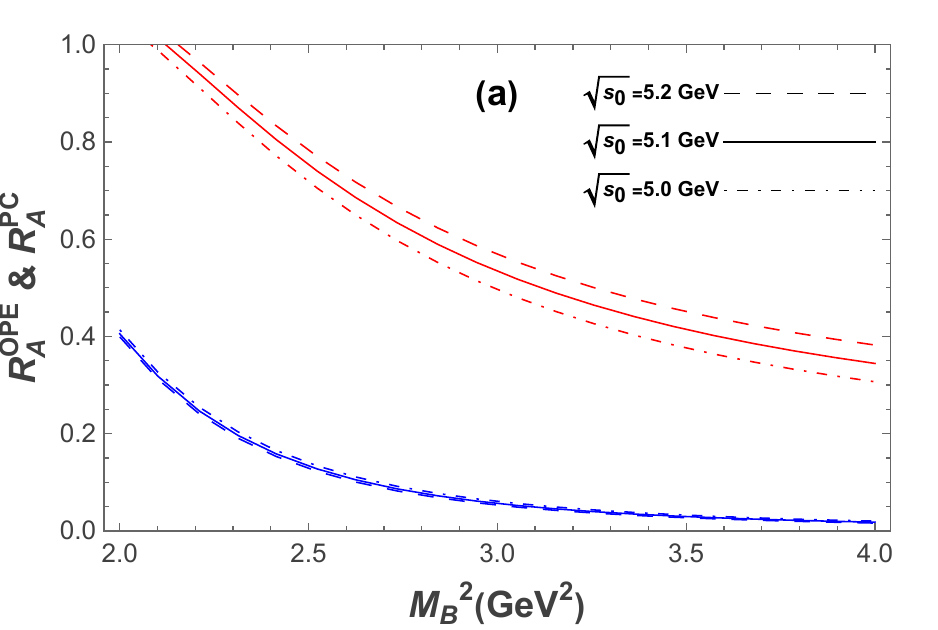}
\includegraphics[width=6.8cm]{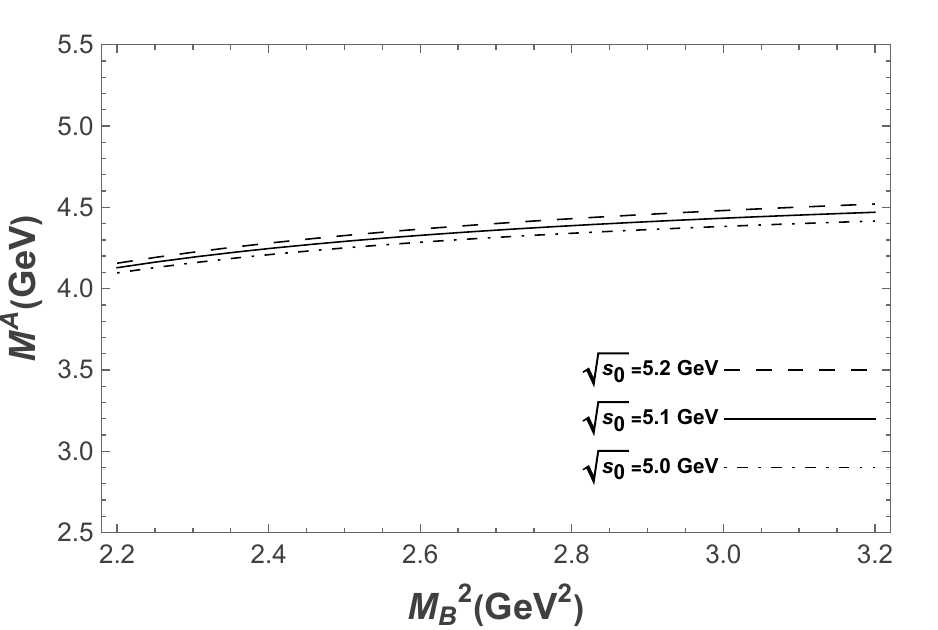}
\caption{The same caption as in Fig.~\ref{figAb0--}, but for the hidden-charm tetraquark state associated with the current $J_A$.} \label{figA0--}
\end{figure}

\begin{figure}[h]
\includegraphics[width=6.8cm]{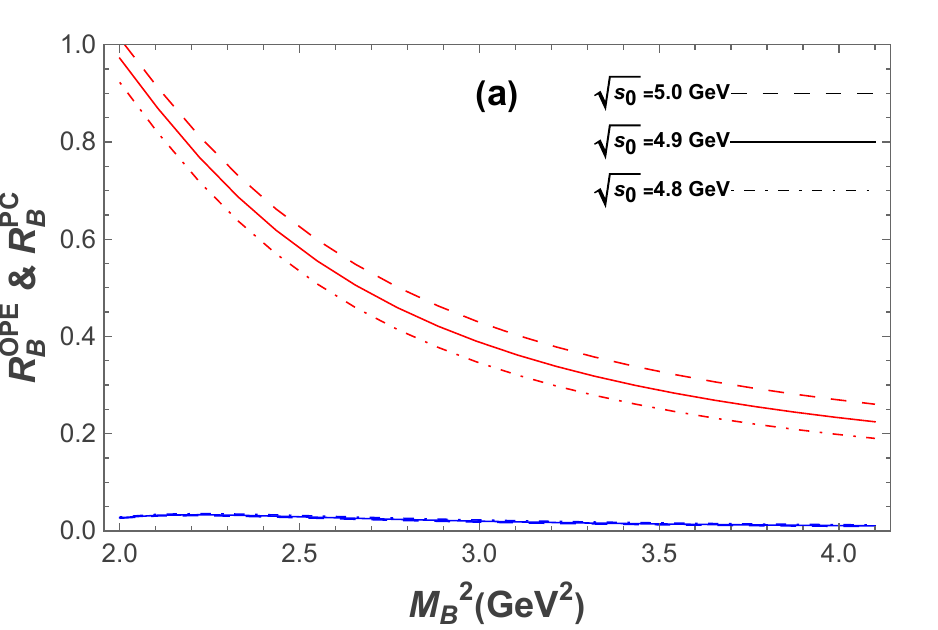}
\includegraphics[width=6.8cm]{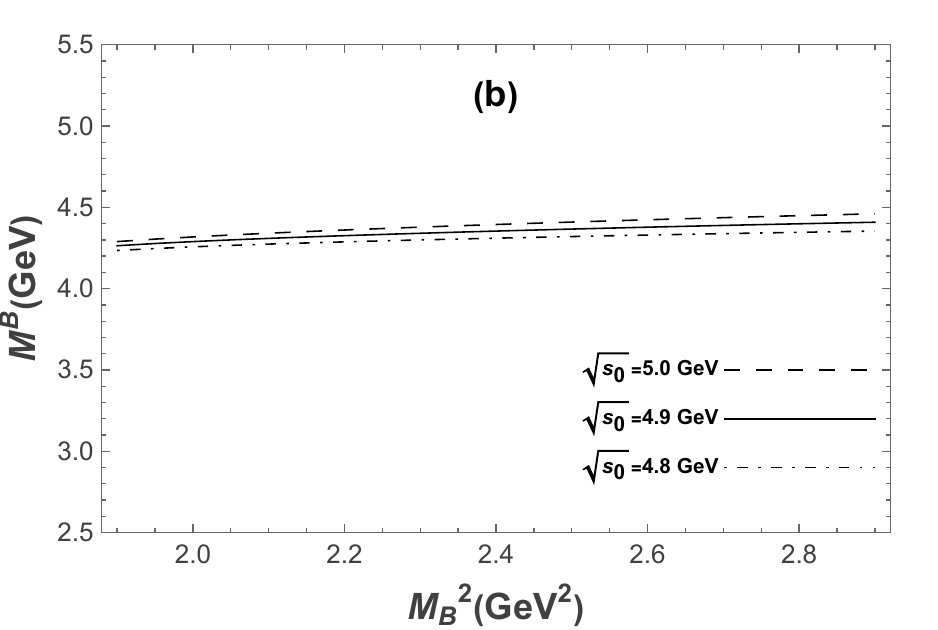}
\caption{The same caption as in Fig.~\ref{figAb0--}, but for the hidden-charm tetraquark state associated with the current $J_B$.} \label{figB0--}
\end{figure}

\begin{figure}[h]
\includegraphics[width=6.8cm]{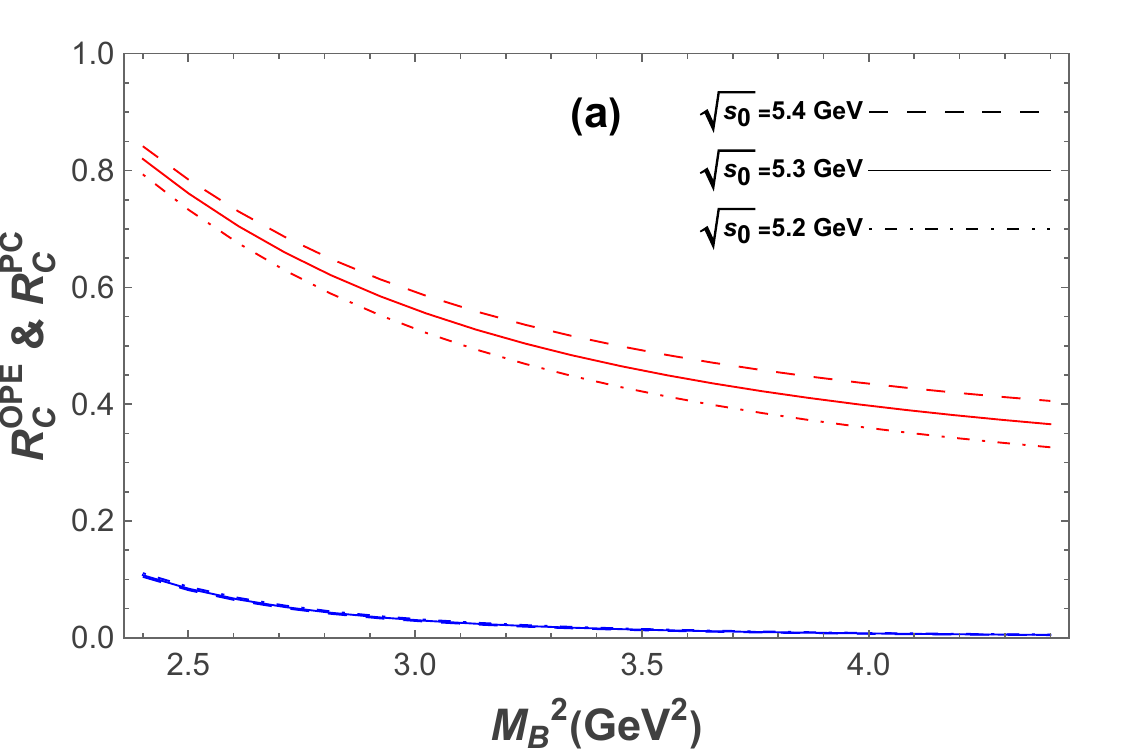}
\includegraphics[width=6.8cm]{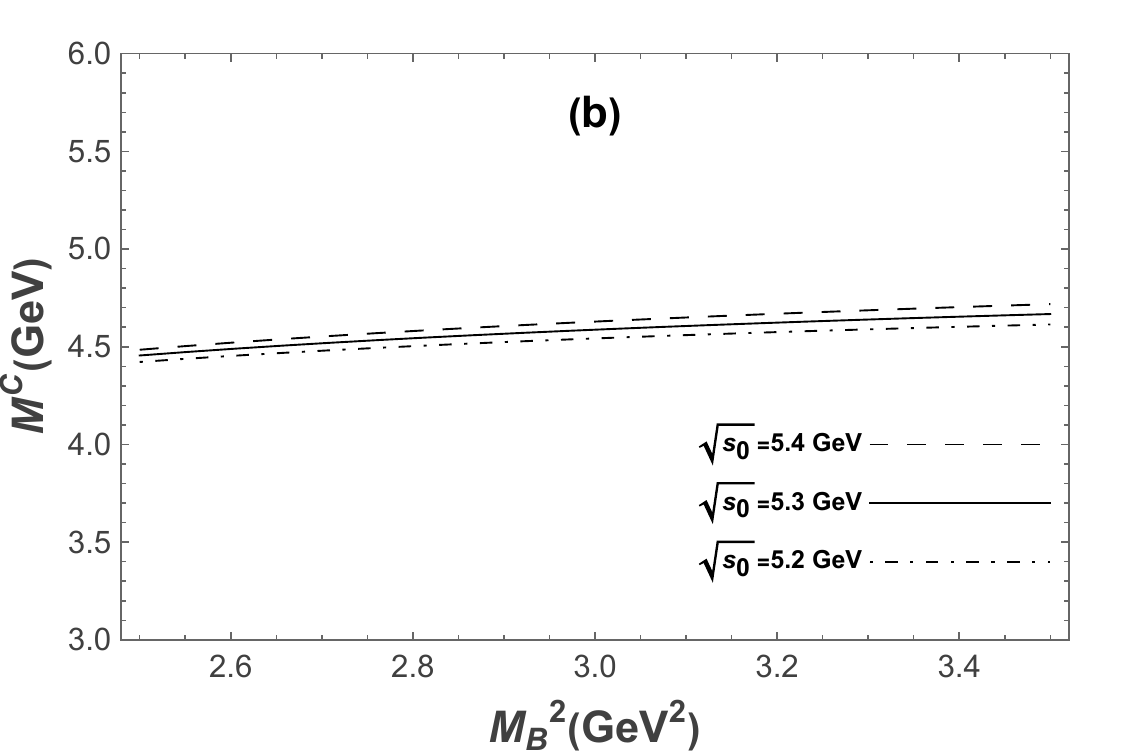}
\caption{The same caption as in Fig.~\ref{figAb0--}, but for the hidden-charm tetraquark state associated with the current $J_C$.} \label{figC0--}
\end{figure}

\begin{figure}[h]
\includegraphics[width=6.8cm]{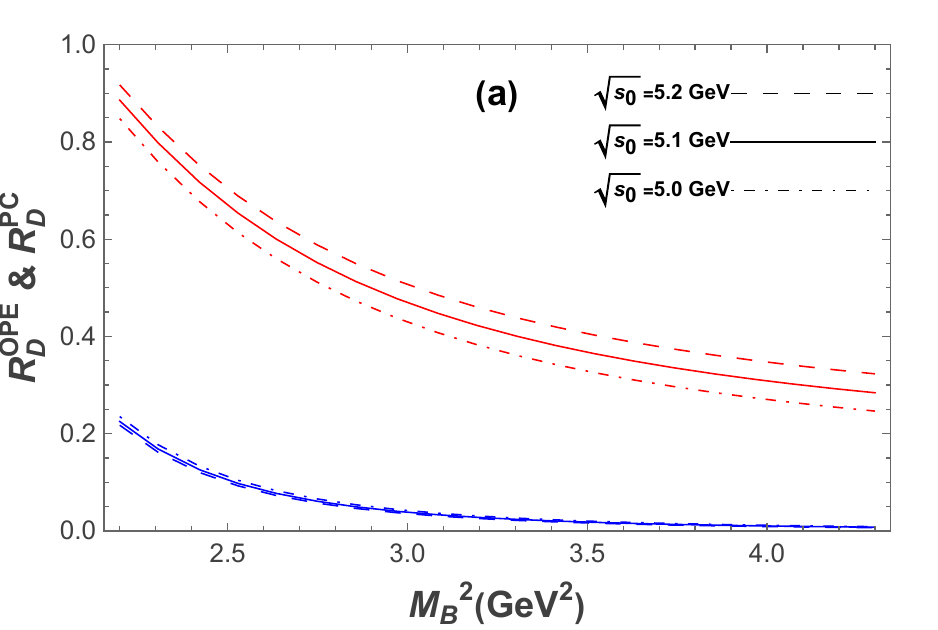}
\includegraphics[width=6.8cm]{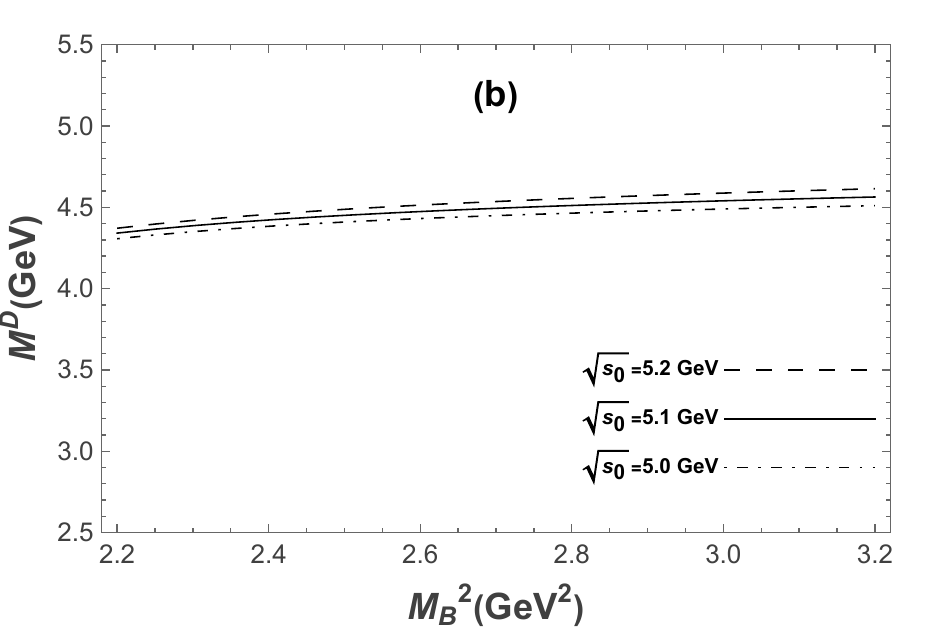}
\caption{The same caption as in Fig.~\ref{figAb0--}, but for the hidden-charm tetraquark state associated with the current $J_D$.} \label{figD0--}
\end{figure}

\begin{figure}[h]
\includegraphics[width=6.8cm]{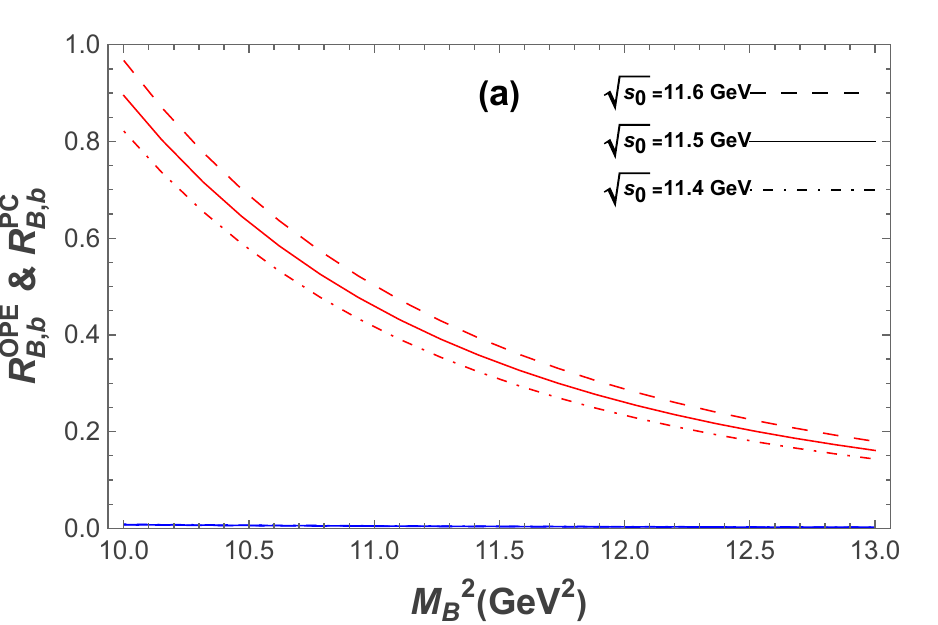}
\includegraphics[width=6.8cm]{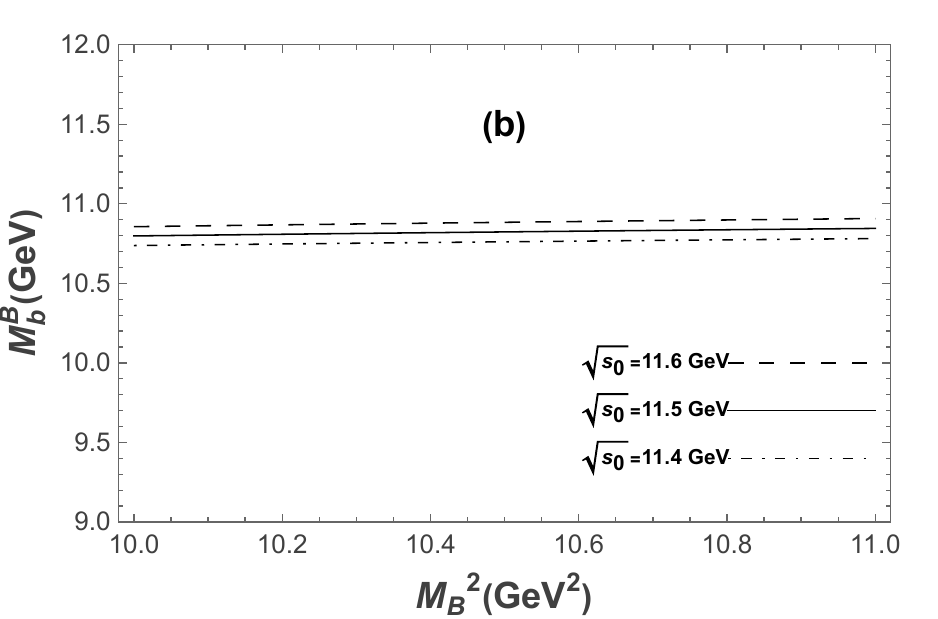}
\caption{The same caption as in Fig.~\ref{figAb0--}, but for the hidden-bottom tetraquark state associated with the current $J_B$.} \label{figBb0--}
\end{figure}

\begin{figure}[h]
\includegraphics[width=6.8cm]{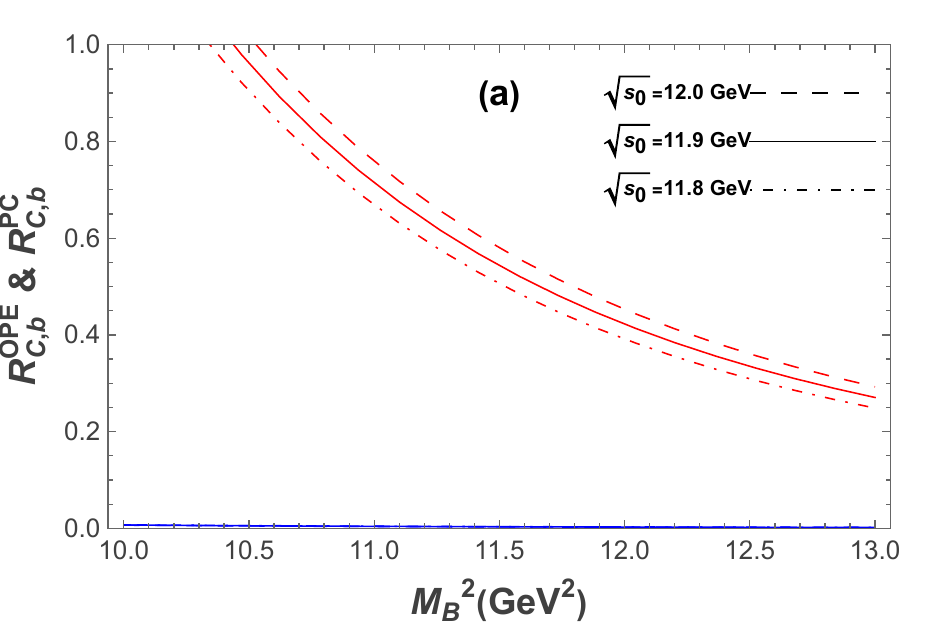}
\includegraphics[width=6.8cm]{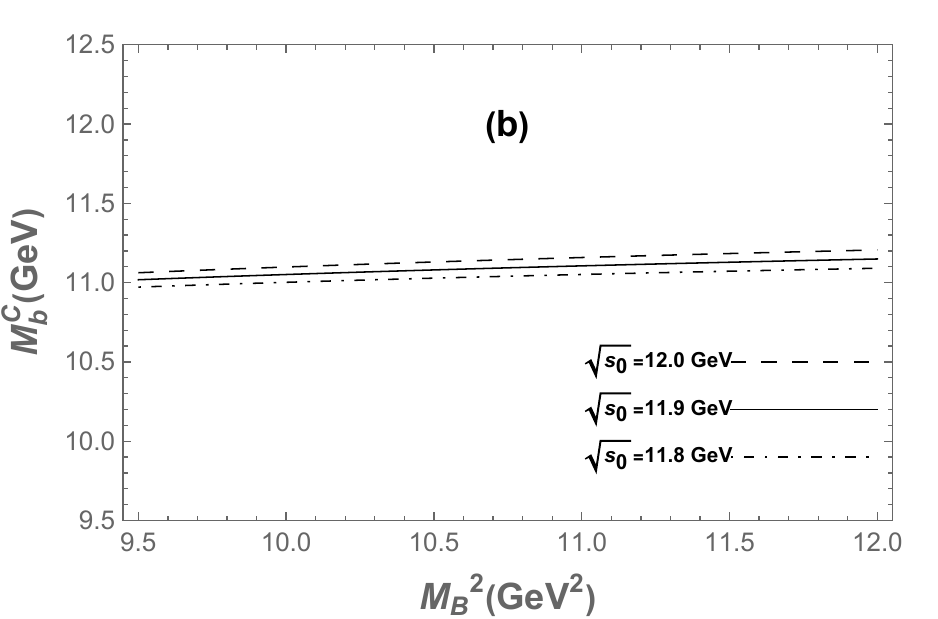}
\caption{The same caption as in Fig.~\ref{figAb0--}, but for the hidden-bottom tetraquark state associated with the current $J_C$.} \label{figCb0--}
\end{figure}

\begin{figure}[h]
\includegraphics[width=6.8cm]{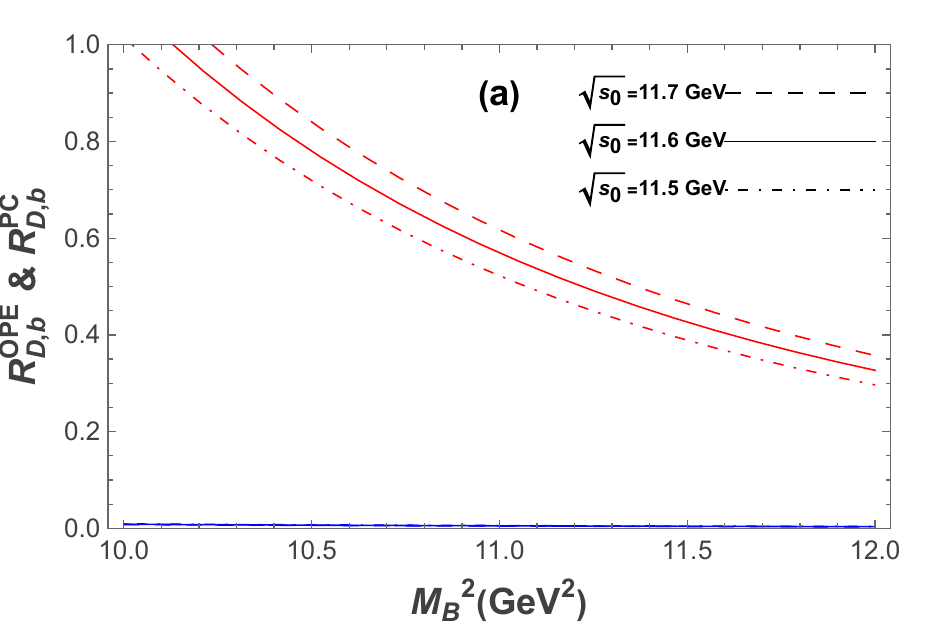}
\includegraphics[width=6.8cm]{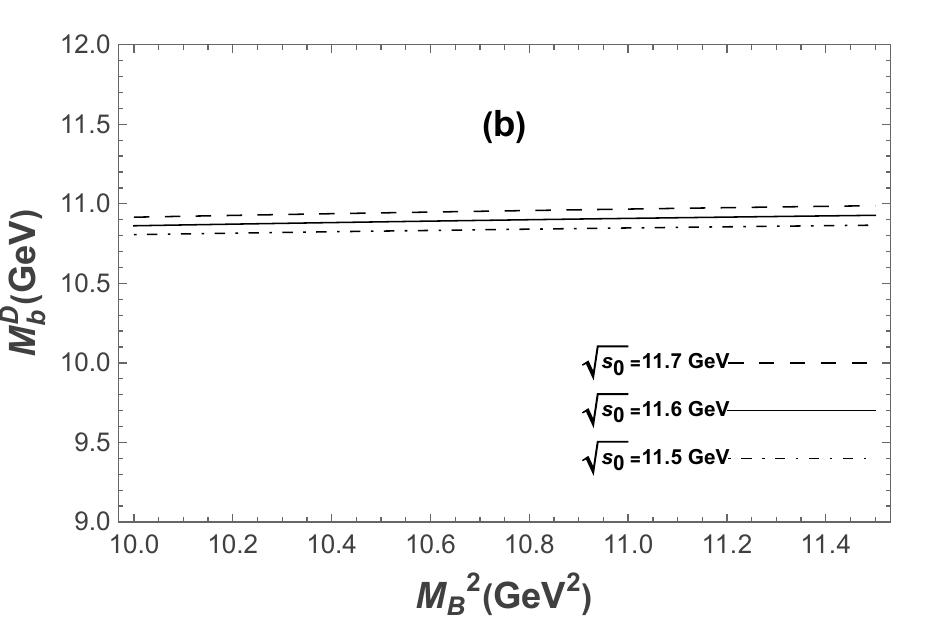}
\caption{The same caption as in Fig.~\ref{figAb0--}, but for the hidden-bottom tetraquark state associated with the current $J_D$.} \label{figDb0--}
\end{figure}

\end{widetext}
\end{document}